\documentclass{egarxiv}
\ConferencePaper

\usepackage[T1]{fontenc}
\usepackage{dfadobe}  

\usepackage{cite}  
\BibtexOrBiblatex
\electronicVersion
\PrintedOrElectronic
\ifpdf \usepackage[pdftex]{graphicx} \pdfcompresslevel=9
\else \usepackage[dvips]{graphicx} \fi

\usepackage{cite}  

\usepackage{booktabs} 
\usepackage{adjustbox}
\usepackage{amsfonts}
\usepackage{xspace}
\usepackage{relsize}
\def\code#1{{{\relsize{-1}\texttt{#1}}}\xspace}

\usepackage{color}

\definecolor{dkgreen}{rgb}{0,0.6,0}
\definecolor{gray}{rgb}{0.5,0.5,0.5}
\definecolor{mauve}{rgb}{0.58,0,0.82}

\definecolor{dkgreen}{rgb}{0,0.6,0}
\definecolor{dkblue}{rgb}{0,0,0.6}
\definecolor{gray}{rgb}{0.5,0.5,0.5}
\definecolor{mauve}{rgb}{0.58,0,0.82}

\usepackage{xcolor}
\definecolor{commentgreen}{RGB}{2,112,10}
\definecolor{eminence}{RGB}{108,48,130}
\definecolor{weborange}{RGB}{255,165,0}
\definecolor{frenchplum}{RGB}{129,20,83}

\usepackage{listings}
\def\code#1{\texttt{#1}}

\title{GPGPU-Parallel Re-indexing of Triangle Meshes
  with Duplicate-Vertex and Unused-Vertex Removal}

\author[I. Wald]{Ingo Wald\\NVIDIA}

\begin{document}

\maketitle
\begin{abstract}We describe a simple yet highly parallel method for
  re-indexing ``indexed'' data sets like triangle meshes or
  unstructured-mesh data sets---which is useful for operations such as
  removing duplicate or un-used vertices, merging different meshes,
  etc. In particlar, our method is parallel and GPU-friendly in the
  sense that it all its steps are either trivially parallel, or use
  GPU-parallel primitives like sorting, prefix-sum; thus making it
  well suited for highly parallel architectures like GPUs.
\end{abstract}

\section{Introduction}

Many applications and algorithms operate on some form of ``indexed''
data structures like triangle or general polygon meshes, tetrahedral
or general unstructured meshes, etc.  In these data structures, a set
of ``vertices'' (usually stored in a single contiguous array) can be
shared among multiple different ``elements'' (such as triangles or
tetrahedra) by having the latter be defined through sets of integer
indices that reference elements in the vertex array.

It is often desirable for such mesh data structures to not contain
either un-used or duplicate vertices; however, many operations that
one might perform on a mesh---such as building a mesh from individual
elements, merging or splitting meshes, computing sub-sets of a mesh,
etc---are most easily implemented in a way that produces just such
vertices.

Thus, it is often required to remove such duplicate and/or un-used
vertices, which if often called ``re-indexing'' or ``re-meshing''.
Conceptually, this is most easily done by starting with a new empty
mesh, adding vertices as required, and keeping track (e.g., using a
\code{std::map}) of which vertices have already been added to the
mesh, and with which index. E.g., for a triangle mesh, this could look
roughly like this:
\begin{verbatim}
    Mesh remesh(Mesh oldMesh)
       Mesh newMesh = empty mesh;
       std::map<vertex,int> alreadyInserted;
       for (triangle in oldMesh)
          for (each vertex v in triangle) 
             if (alreadyInserted
                 .alreadyContains(v)):
                v_idx = alreadyInserted[v];
             else
                v_idx = newMesh.vtx.size()
                newMesh.vtx.push_back(v)
                alreadyInserted[v] = v_idx;
             ....
\end{verbatim}
In particular since the appearence of the Standard Template Library
(STL) this solution is trivially simple to code, and at least for
small meshes works quite well. However, there are  two issues with
this approach: First, it is intrinsically serial, and even on a CPU
can quickly become a bottleneck once meshes become non-trivially
small. And second, this serial nature makes it manifestly unsuitable
to use on GPUs.

In this article, we describe a GPGPU-parallel method for building a
duplicate- and unused vertex-free mesh from a given input mesh. With
``GPGPU''-parallel we mean that our method can be expressed solely
through a sequence of operations that are either trivially
parallelizable (such as setting all elements of an array to a certain
value), or are operations that are readily available in parallel form,
such as parallel sorting or parallel prefix sums in libraries such as
\code{thrust}~\cite{thrust} or Threading Building Blocks
(\code{TBB})~\cite{TBB}. We walk the user through our algorithm using
a simple example, but also provide reference source code for both GPU
(using \code{thrust} and \code{CUDA}~\cite{CUDA}), and CPU (using
TBB).

The resulting algorithm is simple and parallel, works well on both
CPUs and GPUs, and is simple enough for us to be certain that others
must have invented and used it before, many times. However, in absence
of any easily findable yet detailed description of this method we
ourselves have ended up repeatedly re-inventing this method  for
different applications. The goal of this article is to save others
this effort, and to provide an easy-to-find reference description (and
implementation) of this useful method.

\section{The Method}

In this section we will walk the reader through our method in a step
by step fashion; to better illustrate how these steps might affect a
given input mesh we will be illustrating each step using a simple test
case that contains two duplicate and two un-used vertices. In
particlar, we will be using the following triangle mesh with named
vertices \code{A,B,...}, and ``triangles'' of three vertex indices
each:
\begin{verbatim}
    > vtx = {A,B,C',X,D',C",E,F,Y,D"}
    > idx = {(0,1,2)(0,2,4)(5,6,7)(5,7,9)}
\end{verbatim}
Note for that this example it does not
matter whether the vertices are 2, 3, or any other dimensional
vertices (or what other type of data, for that matter); however, we do
assume for this paper that vertices with the same alphabetical letter
are exact duplicates of each other (the primes and quotes just illustrate different
duplicates of a given vertex), and that vertices with different
alphabetical letters are different. We also assume that there is a
obvious way of ordering two vertices as to which one is ``less'' than
another.

In our sample mesh, the vertices \code{X} and \code{Y} are not used by
any triangle; the vertices \code{C} and \code{D} are each stored twice
in different locations---to show how these move during our algorithm
we have intentionally tagged these with either one or two primes;
however, the value of \code{C'} and \code{C''}---respectively
\code{D'} and \code{D''}---are considered to be the same. The actual
four triangles formed by this mesh are, obviously, \code{(A,B,C)},
\code{(A,C,D)}, \code{(C,E,F)}, and \code{(C,F,D)}.

\subsection{Step 1: Replace un-used vertices w/ any used one}

The core goal of our method (which we describe below) is not aimed at
removing \emph{unused} vertices, but only at removing \emph{duplicate}
vertices. However, we can make it solve the problem of unused
vertices, too, by simply replacing each unused vertex with any other
used one (e.g., the first vertex of the first triangle): this
transforms unused vertices into duplicate ones, and though after this
step these now-duplicate vertices are still not referenced by any
triangle they will in the later stages be merged with the used vertex
whose value they were overwritten with, thus effectively removing
them.

To do this replacing of unused vertices with a used one we first
allocate a temporary array \code{isUsed[]} with one bool per vertex,
and initialize this to false for each vertex. Second, we
parallel-iterate over each vertex index used in any of the mesh's
index elements (i.e., each vertex index of each triangle), and mark
that index's corresponding array element as used:
\begin{verbatim}
    parallel_for(each vertex index vtxID)
       isUsed[vtxID] = true;
\end{verbatim}
This operation is obviously parallel and can be trivially implemented
in CUDA, using TBB, or through some \code{thrust::scatter}, etc.  For
our test case, this produces
\begin{verbatim}
    : vtx    = {A,B,C',X,D',C",E,F,Y,D"}
    > isUsed = {1,1,1, 0,1, 1, 1,1,0,1 }
\end{verbatim}
(to improve readability we will, from here on, annotate each step's
\emph{input} arrays with a colon (\code{:}), and outputs from this
step with a greater-than sign (\code{>}).

Finally, we execute another parallel for over all vertices, check if
the vertex is marked as used, and if not, overwrite it with the first
triangle's first vertex: 
\begin{verbatim}
    parallel_for(i = 0 ... numVertices)
       if (!isUsed[i])
          vtx[i] = vtx[triangle[0].vtxIdx[0]]
\end{verbatim}
For our sample mesh, this evaluates to
\begin{verbatim}
    # before overwriting unused vertices
    : vtx    = {A,B,C',X,D',C",E,F,Y,D"}
    : isUsed = {1,1,1, 0,1, 1, 1,1,0,1 }
    # after overwriting unused vertices
    > vtx    = {A,B,C',A',D',C",E,F,A",D"}
\end{verbatim}
which no longer contains the previously unused vertices \code{X} and
\code{Y}, but instead use two new copies of the used vertex \code{A}.

\subsection{Step 2: Make each reference to a duplicated vertex point to the same instance of that vertex}

In this step, we want to ensure that for any duplicate vertex (e.g.,
\code{C} being duplicated into \code{C'}, \code{C''}) all indices
referring to the different copies of this vertex will get replaced
with a single index to only a single one of those copies (all other
copies will remain in the vertex array for now, but have no indices
pointing to them any more). Whether all point to \code{C'} or
\code{C''} does not matter, as long as all point to the same.

To do this, we will first sort all vertices, which will then allow us
to spot duplicate vertices by simply comparing them to their
predecessor. This sorting will change the order of the vertices; so we
first need to ensure that we can later undo this re-ordering. To do
this, we take three steps: First, we create a copy of our (cleaned)
vertex array \code{oldVtx = vtx}. Second, we create an array
\code{orgID[]} with one int per vertex, and fill that with the
identity (for example, using \code{thrust::sequence}):
\begin{verbatim}
    > orgID = { 0,1,2,....}
\end{verbatim}

Using this helper array, we now perform a key-value sort with the
vertex array being the key, and the ID-array being the
value. Key-value sorts are commonly available in different parallel
programming libraries, for example via \code{thrust::key\_value\_sort()}
in thrust, or via \code{tbb::parallel\_sort()} with a
\code{tbb::zip\_iterator} in TBB. For our sample data, this produces
\begin{verbatim}
    # before sorting
    : vtx   = {A,B,C',A,D',C",E,F,A,D"}
    : orgID = {0,1,2, 3,4, 5, 5,6,8,9}
    # after key-value sort
    > vtx   = {A, A, A, B, C",C',D',D",E, F}
    > orgID = {0, 3, 8, 1, 5, 2, 4, 9, 5, 6}
\end{verbatim}
Note the ordering of \code{C'} vs \code{C''} (or \code{D'} vs
\code{D''}) is completely arbitrary, since both \code{C'} and
\code{C''} are the same value - we just tagged one with a prime to
show it has a different index, but they are indistinguishable for the
sort operation, so for an ``unstable'' sort algorithm may end up in
any order.  Since the \code{orgID} array has undergone exactly the
same permutation as the vertices we will later on be able to use this to reverse
this sort operation (Section~\ref{sec:reverse-order}).

Using this sorted array, we can now easily define which vertices are
duplicates of another same vertex with lower index, and which ones are
the respectively first occurrence of a given vertex. To do this we
compute an array \code{nodup[]} (with one bool per vertex), and
execute a parallel kernel that sets each element entry to true if it is either the
first vertex in the list, or different from its predecessor:
\begin{verbatim}
    parallel for i = 0..numVertices
      nodup[i] = (i == 0) || (vtx[i] != vtx[i-1]
\end{verbatim}
For our example:
\begin{verbatim}
    : vtx   = {A, A, A, B, C",C',D',D",E, F}
    > nodup = {1, 0, 0, 1, 1, 0, 1, 0, 1, 1}
\end{verbatim}

Next, we can compute a new array index \code{newIdx} as a postfix sum
over \code{nodup}:
\begin{verbatim}
    : nodup  = {1, 0, 0, 1, 1, 0, 1, 0, 1, 1}
    > newIdx = {1, 1, 1, 2, 3, 3, 4, 4, 5, 6}
\end{verbatim}
then subtract one from each element of this array:
\begin{verbatim}
    > newIdx = {0, 0, 0, 1, 2, 2, 3, 3, 4, 5}
\end{verbatim}
Again, postfix sums are available in common parallel programming
packages, e.g., via \code{thrust::inclusive\_scan}; to then subtract
one we use a trivially simple CUDA kernel\footnote{This subtracting of
  one could also be avoided entirely by setting the first element in
  the \code{nodup} array to 0 instead of 1, but we chose not to do
  this for didactic reasons}. This array now gives, for each vertex, a
new unique array index for an array that would not contain duplicates.

\subsection{Step 3: Computing the New Vertex Array}

In the previous section's \code{newIdx} array, the least index by
design must be the highest vertex index, so the number of unique
elements in our target vertex array is the last element plus one:
\begin{verbatim}
    > newN = newIdx.back()+1 = 6
\end{verbatim}

We can now allocate a new array of vertices with \code{newN} entries;
then, for each index $j<N$ in the original array of vertices, read the
vertex at position \code{j} in the original vertex array, and write it to
position \code{newIdx[j]} in the new vertex array.
\begin{verbatim}
    parallel for i=0..numVertices 
       if nodup[i] // < this is optional
          newVtx[newIdx[i]] = vertex[i]
\end{verbatim}
This operation is identical to a \code{scatter} operation, so where
available can also be implemented as such (e.g., via
\code{thrust::scatter}).

For our sample data, this produces:
\begin{verbatim}
    > newVtx = { A, B, C", D',E, F}
\end{verbatim}
(again, it does not matter which version of \code{C'/C''} or
\code{D'/D''} were written, they have the same value).

\subsection{Step 4: Computing the New Index Array}
\label{sec:reverse-order}

The previous step has computed a new and compact array of vertices
that no longer contains duplicate vertices; however, this---and the
preceding sort operation---have necessarily moved vertices from their
original array locations, so the old vertex indices are no longer
valid, and must be updated.

To do this, we first create an array \code{perm} of one int per
original input vertex, and for each element set \code{perm[orgID[j]]} to
\code{j}.
\begin{verbatim}
    parallel_for(i=0..numVertices)
       perm[orgIdx[i]] = i
\end{verbatim}
(again, this could be done with a \code{thrust::scatter} operation).
After this operation the \code{perm} array effectively describes the
permutation done by sorting. For our example:
\begin{verbatim}
    : orgID = {0, 3, 8, 1, 5, 2, 4, 9, 5, 6}
    > perm  = {0, 3, 5, 1, 6, 4, 8, 9, 2, 7}
\end{verbatim}
With this, we can now update every old index \code{i} to the value
\code{newIdx[perm[i]]}:
\begin{verbatim}
    : idx(old) = {(0,1,2)(0,2,4)(5,6,7)(5,7,9)}
    : perm     = {0, 3, 5, 1, 6, 4, 8, 9, 2, 7}
    : newIdx   = {0, 0, 0, 1, 2, 2, 3, 3, 4, 5}
    > idx(new) = {(0,1,2)(0,2,3)(2,4,5)(2,5,3)}
\end{verbatim}
To explain why this works: The first lookup (\code{i}$\rightarrow$\code{perm[i]})
translates from the original vertex index \code{i} to where this vertex
would have been after sorting; the second (\code{perm[i]}$\rightarrow$\code{newIdx[perm[i]]}) translates from the position after sorting to the
position after compacting (ie, after removing the duplicates)---which
is where the vertex at \code{i} has ended up after both steps.

\subsection{Checking our Example}

To visually show that this is correct for our example, let's first
look at the original mesh with duplicates:
\begin{verbatim}
  : vtx(old) = {A,B,C',X,D',C",E,F,Y,D"}
  : idx(old) = {(0,1,2)(0,2,4)(5,6,7)(5,7,9)}
  > tris(old)= {(A,B,C')(A,C',D')(C",E,F)(C",F,D")
\end{verbatim}
Then, for reference, the new mesh looks as follows:
\begin{verbatim}
  : newVtx = { A, B, C", D',E, F}
  : newIdx = {(0,1,2)(0,2,3)(2,4,5)(2,5,3)}
  > tris   = {(A,B,C)(A,C",D')(C",E,F)(C",F,D')}
\end{verbatim}
These are indeed the same triangles with the same vertices, except for
the arbitrary choice of which copy of a vertex may have been chosen
for duplicate vertices (keep in mind that C' and C''---and D' and
D''---are actually the same vertices, with the prime only used for
illustration of the data movement).

\section{Sample Use Cases}

Though there are probably more uses for this ``re-indexing'' of
meshes, in practice we have found this technique useful in particular
for the following tasks:

\subsection{Removing duplicate vertices}

Sometimes meshes can contain duplicate vertices, either because
whoever created the meshes did not care about removing them, or
because vertices contained some attributes that made them different
(e.g., vertex colors) that the application then discarded. Simply
running the above method will create a new mesh that has the same
triangles, but has all duplicate vertices removed.

\begin{table*}[h]
  \centering
  \begin{tabular}{c|ccccc}
    N &                                8    &     64 &  1024 &  4096 & 8192 \\
    \hline
    mesh: number of quads (in)     &  64    &  4.10K & 1.05M & 16.78M & 67.11M\\
    mesh: number of vertices (in)  & 320    & 20.48K & 5.24M & 83.89M & 335.54M\\
    mesh: number of vertices (out) &  81    &  4.22K & 1.05M & 16.79M & 67.13M\\
    \hline
    time (CPU, serial)             & $<$1ms & 1.1ms  & 854ms &  17.6s & 78s\\
    time (CPU, parallel w/TBB)     & $<$1ms & 1.6ms  &  94ms &   1.5s & 6.2s\\
    time (GPU, thrust+CUDA)        & $<$1ms & $<$1ms & 9.2ms &   92ms & 338ms\\
  \end{tabular}
  \caption{\label{tab:results}Evaluation of our method---comparing
    both scalar CPU reference, our parallel method on CPU, and a CUDA
    implementation thereof---for various different mesh sizes.}
\end{table*}

\subsection{Merging meshes that share (some) vertices}

This is a special case of the previous one: to merge two (or more)
individual meshes, one can in a first step just concatenate the
different meshes' vertex arrays, then increase all of the second
mesh's vertex indices by a value that equals how many vertices the
first mesh originally contained, and append all those increased
indices to the index array, etc. This oeration is cheap and trivially
parallelizable, but ends up with potentially duplicate vertices that
the above method can then be used to remove.

\subsection{``Soup to Mesh'' conversion}

If the input is a set of individual ``fat'' triangles
\code{(v0,v1,v2)}, \code{(v3,v4,v5)}, \code{(v6,v7,v8)} etc, then the
above can be used to create a mesh representation by first creating a
``dummy'' mesh with \code{vtx=\{v0,v1,...\}} and
\code{idx=\{(0,1,2)(3,4,5),(6,7,8)...\}}; then using our algorithm to
remove duplicates---the result is a valid and compact mesh
representation.



\subsection{Computing a subset of a mesh (e.g., splitting a mesh into sub-meshes)}

Some operations require creating a mesh from a subset of triangles of
another mesh, such as the set of all those triangles that share a
given material type\footnote{One common example is the wavefront OBJ
  format, where one typically wants to split the input based on group
  or material type, but where all vertex indices are specified
  relative to a single global vertex array.}. Other examples are
computing a compact mesh of only outer faces of a tetrahedral mesh, or
a new unstructured mesh that only contains the tetrahdra from a
mixed-element mesh, etc.

To create a compact mesh of only a subset of elements, one can simply
create a new mesh that contains a copy of the full vertex array, but
only those elements that one wants in this mesh; the above method will
then remove all unused vertices and return a new, compact mesh with
only used vertices.

\subsection{Reference Code and Performance}

Though we have previously used this method in many different contexts,
for the purpose of some at least rough evaluation we also implemented
some reference code that we have made available via MIT license
under \url{https://github.com/ingowald/sampleCode-parallel-mesh-reindexing}. For
this reference code we chose \code{float2} vertices and \code{int4}
indices because such meshes were considered the easiest to generate
test cases for (see below); adapting this to other
vertex and index formats should be trivial.

To evaluate the impact of parallelization we implemented three
methods: a serial CPU reference method uses a \code{std::map} to track
which vertices have already been added; a parallel version of our
method on the host that uses Threading Building Blocks
(TBB~\cite{TBB}; using the \code{tbb-dev} package on \code{Ubuntu
  20.04}) for parallelization and parallel sorting; and a GPU-parallel
method that uses a combination of thrust~\cite{thrust} (for sorting,
prefix sum, etc) with small CUDA kernels for computing some of the
other values. All three methods have been implemented as we would have
used them in our own codes; careful tuning might give additional
speedups, but this we believe to be true for each one of the three
methods, so their relative performance should be roughly
representative. All experiments are run on a desktop PC with a Intel
Core i7-7820X CPU (8 physical cores at 3.6~GHz/4.3~GHz in
regular/turbo mode). The GPU we used was aN RTX~3090 ``Founders Edition''
GPU with 10,496 CUDA cores at 1.7~GHz, and 24~GB GDDR6X memory.

To evaluate scalability across different mesh sizes we implemented a
generator that would, for a given \code{N}, generate a regular mesh of
$N\times N$ quadrilaterals (i.e., $(N+1)\times(N+1)$ unique vertices) for
which the vertices \emph{are} at the same location as those of its
neighbor quads, but where these vertices are nonetheless
replicated. To also include some unused vertices we also place one
unused vertex into the center of each quad, so every fifth vertex in
the vertex array is unused.

The results of this evaluation---for various values of $N$---are given
in Table~\ref{tab:results}. As can be seen from this, for smaller mesh
sizes there is no major difference between the three methods; however,
for larger meshes parallelization becomes very useful even on the
CPU.

\subsection{Summary and Discussion}

We have described a simple---and in particular, parallel---method for
re-indexing triangle meshes that removes duplicate vertices, and can
also be used for remove unused vertices, for computing compact
sub-sets of larger meshes, etc. In particular, the algorithm is well
suited to GPU execution because all it requires are some trivially
parallel kernels for reading from and writing to arrays, as well as
parallel sort and parallel prefix sum that commonly available
libraries like \code{thrust} already provide.

Though we have illustrated this algorithm only for triangle meshes, it
works exactly the same for any other ``indexed'' representation such
as quad meshes, general polygon meshes, unstructured tetrahedral,
hexahedral, etc meshes, etc. And though its main use case obviously is
to enable fast re-indexing on GPUs (for which a serial solution would
not make any sense whatsoever) at least for non-trivial mesh sizes it
is much more efficient than the serial variant even on the CPUs.

Arguably the biggest shortcoming of this method is that it requires
two copies of both vertex and index array, plus some additional
(though smaller) helper arrays, plus whatever temporary memory the
sorting algorithm may require (which can again be as much as the data
that is being sorted); which for really large meshes may become an
issue.

\nocite{TBB}

\footnotesize
\bibliographystyle{eg-alpha}
\bibliography{article}
\end{document}